\documentclass[twocolumn,epsfig,pre]{revtex4}
\usepackage{graphicx}
\usepackage{amssymb}
\usepackage{amsmath}
\usepackage{hyperref}
\usepackage{latexsym}

\def\be{\begin{equation}}
\def\ee{\end{equation}}
\def\bea{\begin{eqnarray}}
\def\eea{\end{eqnarray}}

\def\e{\epsilon}
\draft

\begin{document}

\title{Bidirectional transport in a multispecies TASEP model}
\author{Sudipto Muhuri$^{1,2}$, Lenin Shagolsem$^{3,4}$ and Madan Rao$^{1,3}$}
\affiliation
{$^1$Raman Research Institute, C.V. Raman Avenue, Sadashivanagar, Bangalore 560080, India\\
$^2$Institute of Physics, Sachivalaya Marg, Bhubaneswar 751005, India\\
$^3$National Centre for Biological Sciences (TIFR), Bellary Road, Bangalore 560065, India\\
$^4$ Leibnitz Institute of Polymer Physics, Dresden, Germany}
\begin{abstract} 
We study a minimal lattice model which describes bidirectional transport of ``particles'' driven along a one dimensional track, as is observed in 
microtubule based, motor protein driven bidirectional transport of cargo vesicles, lipid bodies and organelles such as mitochondria.  This minimal model, a multi-species totally asymmetric exclusion process (TASEP)  with directional switching, can provide a framework for understanding  the interplay between the switching dynamics of individual 
particles and the collective movement of particles in 1-dimension. When switching is  much faster than translocation, the steady state density and current profiles of the particles are  homogeneous in the bulk and are  well described by a Mean-Field (MF) theory, as determined by comparison to a Monte Carlo simulation. In this limit, we construct a non-equilibrium phase diagram. Away from this fast switching regime, the MF theory fails, although the average bulk density profile still remains homogeneous. We study the steady state behaviour as a function of the ratio of the translocation and net switching rates, $Q$, and find a unique first-order phase transition at a finite $Q$ associated with a discontinuous change of the bulk density. When the switching rate is decreased further (keeping translocation rate fixed), the system approaches a jammed phase with a net current that tends to zero as $J\sim 1/Q$.

\end{abstract}

\maketitle
\section{Introduction}

Intracellular transport of molecules or cargo vesicles often proceed bidirectionally along one-dimensional cytoskeletal filaments such as microtubules. The most well studied of these are the dynamics of pigment granules in melanophores \cite{borisy}, mitochondria in the axons of neuronal cells \cite{hollenbeck,welte}, endosomes in {\it Dictyostelium} cells \cite{roop} and lipid-droplets in {\it Drosophila} embryos\cite{lipo,grosslipid}. Bidirectionality is  achieved by binding to oppositely directed motor proteins such as kinesin and dynein working in a regulated manner \cite{cell,howard}.
Typically intracellular transport along a single filament involves many cargo vesicles /organelles, having  a distribution of sizes. In order to have unimpeded and efficient movement of the cargo, the ``single-particle'' switching mechanism must be coordinated with the collective ``many-particle'' behaviour.
This provides the motivation for the study of minimal models of bidirectional transport that highlight this interplay between individual 
switching and collective movement of motile elements.

With this in mind, we extend the popular one-dimensional (1d) driven diffusive models to describe bidirectional movement. In spite of their inherent simplicity, driven diffusive systems exhibit a rich phase diagram characterized by novel non-equilibrium steady states with finite macroscopic  current \cite{schutzrev,nonequi2,privman,barma5} and boundary induced phase transitions \cite{schutzrev,freylet,santen,kolo}. A particular example of such models is the Totally Asymmetric Exclusion Process (TASEP), in which the `particles' hop unidirectionally with a single rate and interact via hard-core repulsion.
Introduction of oppositely directed species in the TASEP model would inevitably lead to jamming. A natural way to avoid this would be to have 
stochastic (or regulated) directional switching of the oppositely directed species.
Here we study a two species TASEP model for driven transport where the two species are oppositely directed and can interconvert stochastically leading to directional switching. This minimal model shows bidirectional transport and a variety of nonequilibrium phase transitions. We focus our attention on the situation where the translocating particles remain attached to the 1-d filament, the conserved model,
first introduced in \cite{thesis}, in contrast to the situation studied in \cite{ignaepl,ignapre}, where translocating particles are subject to Langmuir (un)binding kinetics leading to particle nonconservation in the bulk. 

In Section~\ref{sec:model}, we describe the model and study it using a continuum mean-field (MF) approach in the limit for which the dynamics of switching is much faster than the translocation of the individual species. A comparison of the mean field results with Monte-Carlo simulations show very good agreement for the density and current profile in the bulk. We also find that the steady state solution for average density and current is always homogeneous. In Section~\ref{sec:phases} we use a domain wall analysis \cite{kolo} to study perturbations of the steady states and construct a nonequilibrium phase diagram. The diverse nature of phases are analyzed in terms of variations of switching rate. The model also exhibits switching rate  induced current reversal.
 Away from this fast switching regime, MF theory fails. In  Section~\ref{sec:MFbreakdown}, we use numerical simulations to systematically study departures from this fast switching limit. We study it in terms of the ratio of the translocation and net switching rates, $Q$. We find a first-order phase transition, associated with a discontinuous change of bulk density on variation of $Q$. For a fixed translocation rate when $Q$ is increased further, the system approaches a jammed phase with a net current that approaches zero as $J\sim 1/Q$ and a density profile that is independent of the boundary (un)loading rates.  Finally in Section~\ref{sec:conclusions} we discuss the possible ramifications of these results on filament based intracellular transport.


\section{Two species model}
\label{sec:model}
Cargo vesicles bound to multiple $+/-$ end directed motors moving along a microtubule, either make steps towards the $+/-$ end or stop, as a result of the regulated switching of the bound motors.  Thus, transitions in the cargo vesicle through `chemical space' (defined by the states of the multiple motors) gives rise to movement in physical space 
(a translocation towards the $+/-$ end or stationary).
For a minimal model, we will restrict the instantaneous state of a `particle' to be either in a $(+)$-directed state or $(-)$-directed state, with fixed switching rates between them \cite{gross}. 
Switching can be a result of an  internal regulatory mechanism attributed to self regulation by motors or external regulation \cite{welte}. 
The particle moves  along the 1-dimensional filament of length $L$, represented by a lattice of $N$ sites, labeled ~$i = 0, \dots,N-1$, with lattice spacing 
$\epsilon = L/N$. The sites $i = 0$ and $i = N-1$ define the left and the right boundary, respectively. 
Each site $i$ is occupied by either $(+), (-) $ or a vacancy $(0)$.

The $(+)/(-)$ hop to the (right/left) with the same rate $a$, if and only if the adjacent site is vacant. The rate of switching $k_{+} (k_{-})$ corresponds to interconversion rate from a right(left) moving state to left(moving) moving states. The $(+)$ enters the left boundary with a rate $\alpha_{+}$ if that site is vacant, and leaves the right boundary with a rate $\beta_{+}$. The $(-)$ species enter the right boundary with a rate $\alpha_{-}$ if that site is vacant, and leaves the left boundary with a rate $\beta_{-}$. Figure 1 is a schematic of the various dynamic processes and the corresponding rates.

\begin{figure}[h]
\centering
\includegraphics[width=2.5in,angle=-90]{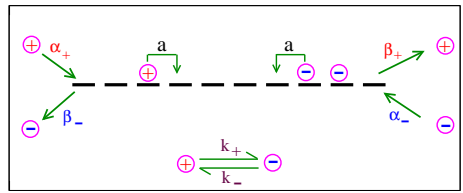}
\begin{center}
\caption{Dynamical rules for the minimal model involving translocation, switching and entry/exit. The reversible switching rates between $(+)$ and $(-)$ are also shown.}
\end{center}
\end{figure}

The discrete equations motion for the left and right moving vesicles at site $i$ are expressed in terms  of  occupation number operators, 
\begin{eqnarray}
\frac{dn_i^{\pm}}{dt} &=&a\left[ n_{~i\mp1}^{\pm}(1 - n_{i}^{+} -n_{i}^{-}) - n_{i}^{\pm}(1 - n_{~i\pm1}^{+} -n_{~i\pm1}^{-})\right] \nonumber\\
&-& k_{\pm}n_{i}^{\pm}+ k_{\mp}n_{i}^{~\mp}
\end{eqnarray}
for $i=1,\dots ,N-2$. The terms on the right hand side  have their usual interpretation of gain and loss terms at each site $i$, arising from translocation and  switching processes. The corresponding current for the  left and right moving particles  can be expressed as 
\begin{equation}
J^{\pm}_{i} = an_{i}^{\pm}(1 - n_{~i\pm1}^{+} -n_{~i\pm1}^{-})
\end{equation}
The vacancy occupation number $n_{i}^0$ is determined by the conservation law $n_{i}^{0} + n_{i}^{+} + n_{i}^{-} = 1 $.

The boundary conditions can be expressed in terms of the instantaneous vacancy number, $n_{i}^{0}$.
At the left boundary site, $i = 0$, a vacancy enters, when either a $(-)$ particle leaves the left boundary or a $(+)$ particle hops to the neighbouring site in the bulk. Similarly a vacancy leaves the boundary site at the left boundary when either a $(+)$ enters the boundary site, $i = 0$ or a $(-)$ particle enters this site the next neighbouring site in the bulk, $i = 1$. Accordingly,
\begin{equation}
\frac{dn_0^{0}}{dt}  =  an_{0}^{+} n_1^0 + \beta_{-}n_{0}^{-} -\alpha_{+}n_0^0 -an_{1}^{-} n_0^0
\label{eq:n0}
\end{equation}
Similarly the evolution of vacancies at the right  boundary reads
\begin{equation} 
\frac{dn_{N-1}^{0}}{dt} = an_{N-1}^{+} n_{N-2}^{0} + \beta_{+}n_{N-1}^{+} -\alpha_{-} n_{N-1}^{0}-  an_{N-2}^{+} n_{N-1}^{0}
\label{eq:n1}
\end{equation}

If one averages the occupation number of each species over the events that may occur between $t$ and $t + dt$ and over all histories up to time $t$, then the corresponding equations of motion for the expectation value of the occupation numbers are given by,
\begin{eqnarray}
\frac{d\langle n_i^{\pm}\rangle}{dt} &=&a[\langle n_{~i\mp1}^{\pm}n_{i}^{0}\rangle - \langle n_{i}^{\pm}n_{i}^{0}\rangle]- k_{\pm}\langle n_{i}^{\pm}\rangle + k_{\mp}\langle n_{i}^{~\mp}\rangle
\end{eqnarray}
For the boundary at $i = 0$,
\begin{equation}
\label{eq:bou1}
\frac{d\langle n_0^{0}\rangle}{dt} = a\langle n_{0}^{+} n_1^0\rangle + \beta_{-}\langle n_{0}^{-}\rangle  -\alpha_{+}\langle n_0^0\rangle -a\langle n_{1}^{-} n_0^0\rangle
\end{equation}
For the boundary at $i = N-1$,
\begin{eqnarray} 
\label{eq:bou2}
\frac{d\langle n_{N-1}^{0}\rangle}{dt} &=& a\langle n_{N-1}^{+} n_{N-2}^{0}\rangle + \beta_{+}\langle n_{N-1}^{+}\rangle -\alpha_{-} \langle n_{N-1}^{0}\rangle \nonumber\\&-&  a\langle n_{N-2}^{+} n_{N-1}^{0}\rangle 
\end{eqnarray}
 To determine the expectation value of occupation number one requires the knowledge of all higher order correlations which makes it analytically intractable \cite{privman}. In specific cases like TASEP, owing to special conservation laws, this moment hierarchy can be handled exactly using matrix methods \cite{asepexact1,asepexact2}. In situations where the fluctuations of the number densities are smaller than mean density, a mean field approach works remarkably well \cite{privman,kolo}. We will see that mean field approach works very well when the process of interconversion is much faster than the translational hopping on the lattice, i.e., $k_{+},k_{-} \gg a$ (referred to as the {\it fast} switching limit). However away from this limit, the mean field approach fails, as seen by comparing with results from Monte-Carlo simulations.

\subsection{Mean field and Continuum limit}
Defining $\rho_i^{+} \equiv n_{i}^{+}$ and $\rho_i^{-} \equiv n_{i}^{-}$, the mean field approximation amounts to factorizing all the two-point correlators arising out of the different combinations of $n_{i}^{+}, n_{i}^{-}$ as a product of their averages, 
\begin{eqnarray}
\langle n_{i}^{\pm}n_{i+1}^{\pm}\rangle &=& \langle n_{i}^{\pm}\rangle \langle n_{i+1}^{\pm}\rangle = \rho_i^{\pm}\rho_{i+1}^{\pm}\nonumber\\
\end{eqnarray}
After taking the expectation values of the operator equations, we go over to a continuum description. We normalize the total length of the lattice, L to 1 and let $N\rightarrow \infty$ so that the lattice spacing $\epsilon = \frac{L}{N}\rightarrow 0$ in the thermodynamic limit.
 The rescaled position variables are defined as  $x = i/N-1,0 \leq x \leq 1$
To go from discrete number representation to continuum density, we Taylor expand,
\begin{eqnarray}
\langle~n_{i}^{\pm}~\rangle &=& \rho_{\pm}(x)\nonumber \\
\langle~n_{i+1}^{\pm}~\rangle &=& \rho_{\pm}(x) + \epsilon\rho_{\pm}^{'}(x) + \frac{\epsilon^{2}}{2}\rho_{\pm}^{''}(x)+ \ldots \nonumber \\
\langle~n_{i-1}^{\pm}~\rangle &=& \rho_{\pm}(x) - \epsilon\rho_{\pm}^{'}(x) + \frac{\epsilon^{2}}{2}\rho_{\pm}^{''}(x) + \ldots
\end{eqnarray}
It useful to re-express the two species densities $\rho_{+}(x)$ and  $\rho_{-}(x)$ in terms of the total cargo density and the relative concentration of the different species,
$\rho = \rho_{+} + \rho_{-}$ and  $ \phi = \rho_{+} - \rho_{-}$. To leading order  in $\epsilon$, the evolution equations then read as,
\begin{eqnarray}
\label{eq:mf_rho}
 \frac{\partial \rho}{\partial t }&=& -\e a\frac{\partial}{\partial x}\left[\phi(1 - \rho)\right]  \\
\frac{\partial \phi}{\partial t}&=& -k_{+}(\rho + \phi) + k_{-}(\rho- \phi) -\e a\frac{\partial}{\partial x}\left[\rho(1 - \rho)\right] 
\label{eq:mf_phi}
\end{eqnarray}

At  steady state, Eq.(\ref{eq:mf_rho}) implies that $\phi(1-\rho)$ is constant in space. We note that $-a\rho(1-\rho)$ is the expression for vacancy current. At steady state, in the absence of any {\it sink} and {\it source} terms in the bulk, the conservation of vacancies in the bulk implies that the vacancy current is spatially homogeneous. This 
implies that both $\phi$ and $\rho$ are individually spatially homogeneous at steady state. 

In the thermodynamic limit of $\epsilon\rightarrow 0$, following Eq.(\ref{eq:mf_phi}), the two densities become proportional to each other,
\begin{equation}
\phi = \frac{(k_{-} - k_{+})\rho}{(k_{+} + k_{-})}=\frac{1-K_+}{W_+}\rho(x)
\label{eq:ss_1}
\end{equation}
where we  have introduced the effective ratio  of interconversion rates   $K_{+} = \frac{k_{+}}{k_{-}}$ and $W_{+} = 1 + K_{+}$. From this relation we obtain the corresponding  relation between the local occupancy of each species   
\begin{equation}
\rho_{+}(x) = K_{+}\rho_{-}(x)
\label{eq:ratio}
\end{equation}
Using eq.(\ref{eq:ratio}), 
the corresponding expressions for the steady state currents for the individual species are,
\begin{eqnarray}
\label{eq:currentexpression}
J_{+}&=&~a~\rho_{+}(1- W_{+}\rho_{+})\\
J_{-}&=&~-a~\rho_{-}(1- W_{+}\rho_{+}) 
\end{eqnarray} 
\indent
Eq.(\ref{eq:mf_rho}) also implies that the currents $J_{+}, J_{-}$ are constant in the bulk with no spatial variation. Further,  they are related to each other by, 
\begin{equation}
J_{-} = - K_{+}J_{+}\nonumber
\end{equation} 
The total particle current in the bulk reduces to,  
\begin{equation}
J(x) = J_{+}(x) + J_{-}(x)=(1-K_+)J_+(x)
\end{equation}
which explicitly shows that switching dynamics due to regulatory mechanism allows for current reversal.   For  $0 < K_{+} < 1$, the overall flux of vesicles is from left to right while for $K_{+} > 1$, the overall flux is from right to left. 

\begin{figure}
\begin{center}
\includegraphics[width=2.5in,angle=-90]{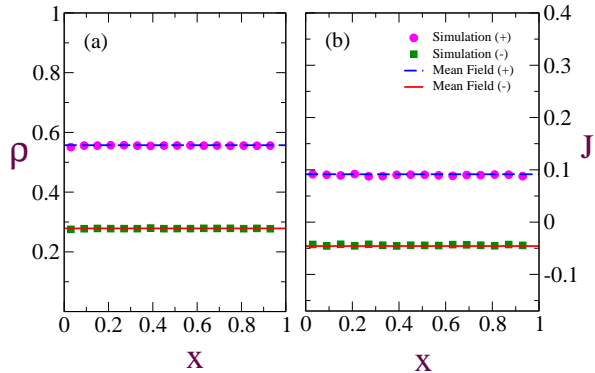}
\caption{Comparison of the steady state density and current profiles obtained by Monte Carlo with the Mean Field theory in the {\it fast} switching limit, shows very good agreement between Monte- Carlo simulation and MF results. These steady state profiles are for the  High Density (HD) phase. (a) Profile for Density of (+) and (-) (b) Normalized current: $J = J_{\pm}/a$. Here $\alpha_{+}/a = 0.3 $, $\beta_{+}/a = 0.2$, $\alpha_{-}/a = 0.4$, $\beta_{-}/a = 0.1$ and $K_{+} = 0.5$. Monte Carlo simulation is done for a system size $N = 500$.}
\label{fig2}
\end{center}
\end{figure}

\subsection{Boundary Condition}

The steady state condition at the boundaries, is that the entry and exit rates of vacancies should match. In the  thermodynamic limit,  the  entry and exit rates of the vacancies at site $i = 0$ read, 
\begin{eqnarray}
R_{ent} &=& \rho_{+}(0)\left[1 - \rho_{+}(0) - \rho_{-}(0)\right] + \beta_{-}\rho_{-}(0) \nonumber\\
R_{ext} &=& [\alpha_{+} + \rho_{-}(0)]\left[1 - \rho_{+}(0) - \rho_{-}(0)\right] 
\end{eqnarray}
This also follows as a consequence of Eq.(\ref{eq:bou1}) and Eq.(\ref{eq:bou2}) after appropriately invoking the Mean-Field limit and then taking the continuum limit.
At steady state $R_{ent}=R_{ext}$. Using the fact that $\rho_{+} = K_{+}\rho_{-}$ we obtain, 
\begin{equation}
\rho_{+}(0) = \frac{M \mp \left[M^{2} -4\alpha_{+}W_{+}\left(1 - K_{+}\right)\right]^{\frac{1}{2}}}{2W_{+}\left(1 -K_{+}\right)}
\label{eq:boundary1}
\end{equation}
where $M = \alpha_{+}W_{+} + \beta_{-}K_{+} + 1 - K_{+}$. Similarly,  at the right boundary of the filament, $x = 1$,
\begin{equation}
\rho_{+}(1) = \frac{P + \left[P^{2} +4\alpha_{-}W_{+}\left(1 -K_{+}\right)\right]^{\frac{1}{2}}}{2W_{+}\left(1 -K_{+}\right)}
\label{eq:boundary2}
\end{equation}
where $P = -\alpha_{-}W_{+} - \beta_{+} + 1 - K_{+}$.


\section{Phases and Domain wall analysis}
\label{sec:phases}

In the limit of $\epsilon \rightarrow 0$, both the boundary conditions cannot be satisfied simultaneously. Therefore either the steady state solution satisfies the boundary condition at the left ($x = 0$) or it satisfies the boundary condition at the right ($x = 1$) and both of these correspond to the two possible stationary states and correspondingly two different phases. 

 In addition, there is a one more phase, the maximal current phase. It corresponds to a situation where the stationary state is independent of the entry and exit rates of particles at the boundaries and the state is determined by maximizing the current in the lattice. Maximizing the expression for currents in eq.(\ref{eq:currentexpression}), the density and current in this phase are,
\begin{eqnarray}
&&\rho_{+} =\frac{1}{2W_{+}} \nonumber\\
&& J_{+} = \frac{1}{4W_{+}} 
\end{eqnarray}\\

Thus, similar to TASEP, there are three possible phases : Low density (LD) phase corresponding to stationary state satisfying the left boundary condition at $x = 0$, a  High density (HD) phase corresponding to stationary state satisfying the right boundary condition at $x =1$ and a maximal current (MC) phase corresponding to the stationary state which is independent of the boundary conditions.

In order to {\it construct} the phase diagram, we need a selection criterion for a given phase. A reasonable selection criterion is provided by the domain wall analysis \cite{kolo}. A domain wall is initially constructed as an interfacial region separating two distinct stationary states. If  this configuration is unstable, then the domain wall will move to totally eliminate the unstable state. At phase coexistence, the domain wall velocity will be zero. 

The expression for the velocity of the domain wall is 
\begin{equation}
V_{d} = \frac{J_{+}^R - J_{+}^{L}}{\rho_{+}^{R}- \rho_{+}^{L}} ,
\label{eq:vd}
\end{equation} 
where $J_{+}^R, \rho_{+}^R$ are the density and current to the right of the domain wall and  $J_{+}^L, \rho_{+}^L$ are the density and current to the left of the domain wall. 


\begin{figure}
\begin{center}
\includegraphics[width=2.5in,angle=-90]{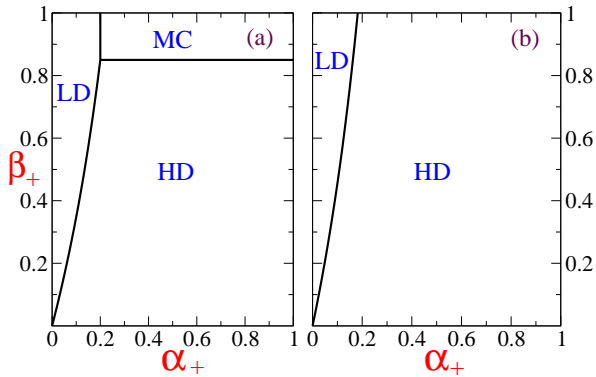}
\caption{Phase diagram in the $\alpha_{+}-\beta_{+}$ plane, for fixed values $K_{+} = 0.5$, $\beta_{-} = 0.1$. (a) $\alpha_{-} = 0.4$ and (b) $\alpha_{-} = 0.6$.  Beyond the multicritical point at $\alpha_{-}^* = 0.5$, there is no MC phase, like in case of (b).}
\label{fig3}
\end{center}
\end{figure}

To obtain the LD-MC phase boundary, we compute the velocity of the LD-MC domain wall. Taking $\rho^{+}_{R}$ = $\rho^{+}_{MC}$ and $\rho^{+}_{L}$ = $\rho^{+}_{LD}$, and putting $V_d = 0$ in eq.(\ref{eq:vd}), we obtain the equation for the LD-MC phase boundary, 
\begin{equation}
\alpha_{+} = \frac{1}{2~W_{+}}\left( 2\beta_{-}K_{+} + 1 - K_{+}\right),
\end{equation}
\noindent
which in the $\alpha_{+} - \beta_{+}$ plane, is a straight line parallel to $\alpha_{+}$ axis. To obtain the MC-HD phase boundary, we compute the velocity of the MC-HD domain wall. Taking $\rho^{+}_{L}$ = $\rho^{+}_{MC}$ and $\rho^{+}_{R}$ = $\rho^{+}_{HD}$, and putting $V_d = 0$, we obtain the equation of the MC-HD phase boundary, 
\begin{equation}
\beta_{+} = \frac{1}{2}\left( 1 + 2\alpha_{-}W_{+} - {K_+}\right),
\end{equation}
which is a straight line parallel to $\beta_{+}$ axis. We see that beyond a critical value of $\alpha_{-}$ (a multicritical point ), the maximal current phase does not occur in physical range of entry and exit rates. Thus the high density phase and the low density phase are the only two phases which can occur in such a situation.
Finally, the LD-HD phase boundary is given by,
\begin{equation}
\beta_{+} = (1-K_{+} -\alpha_{-}W_{+}) + \frac{4 \alpha_{-}(1-K_{+}^{2}) - Z^{2}}{2Z}
\label{eq:betaplus_1}
\end{equation}
where, $Z = 2[\frac{1}{W_+} - \rho_{+}(0)](1-K_{+}^{2})$. 
Note that unlike TASEP, the phase boundary separating the low and high density phases is not a straight line. Fig.\,3 shows the mean field phase diagrams and effect of control parameter in expelling the maximal current phase.

\section{Break down of mean field and approach to jamming}
\label{sec:MFbreakdown}

\begin{figure}
\begin{center}
\includegraphics[width=2.5in,angle=-90]{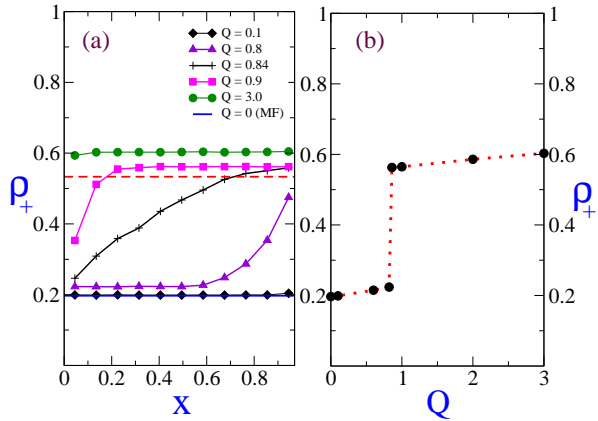}
\caption{(a) Density profile of $\rho_{+}$ as a function of $Q = \frac{a}{k_{+} + k_{-}}$, the ratio of translocation and switching rate, which measures the deviation from the {\it fast} switching limit. The dashed line is the MF prediction for density in HD phase. (b) Plot of the corresponding discontinuous change of bulk density at $Q = 0.84$, a first order transition from a LD phase to a HD phase. Here $\alpha_{+}/a = 0.2 $, $\beta_{+}/a = 0.5$, $\alpha_{-}/a = 0.4$, $\beta_{-}/a = 0.8$ and $K_{+} = 0.25$. Monte Carlo simulation is done for a system size $N = 1000$.}
\label{fig4}
\end{center}
\end{figure}

\begin{figure}
\begin{center}
\includegraphics[width=2.5in,angle=-90]{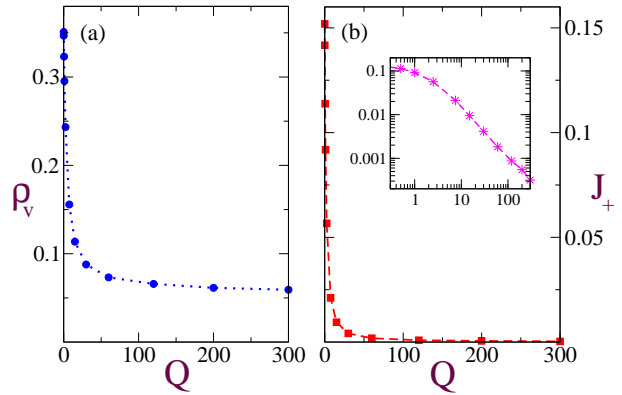}
\caption{(a) Plot of vacancy density as a function of $Q$ is shown. (b) Variation of the current of the $(+)$ species, $J_{+}$ as a function of $Q$ is shown. The current decreases with increasing $Q$, diverging strongly from mean field predictions and asymptotically approaching zero for $Q \rightarrow \infty$ .  Inset figure shows the  plot of $\log Q$ vs $\log J_{+}$ for the same data as in (b). A straight line fit to the inset plot for data points for which $Q\geq 2$ shows that $J \sim Q^{\nu}$, where $\nu \approx -1.097$. Here $\alpha_{+}/a = 0.3 $, $\beta_{+}/a = 0.5$, $\alpha_{-}/a = 0.4$, $\beta_{-}/a = 0.1$ and $K_{+} = 0.5$. Monte Carlo simulation is done for a system size $N = 400$.}
\label{fig5}
\end{center}
\end{figure}


So far our study has been restricted to the fast switching regime, where the rates of interconversion are faster than translocation. In order to study deviations away from this particular limit, we define a dimensionless parameter $Q = \frac{a}{k_{+} + k_{-}}$, which is the ratio of the translocation rate and the net switching rate. Thus {\it fast} switching corresponds to the limit of $Q=0$.

Most realistic cellular situations are far from this fast-switching limit. To see this, 
we estimate the typical values of $Q$ for  transport of lipid-droplets in wild type {\it Drosophila} embryos \cite{grosslipid}. The average run velocity of the lipid droplet, before a switching or pause event, is measured to be $\sim 0.2 \mu m s^{-1}$ for short runs $(0.03-0.1~ \mu m)$ and $\sim 0.5 \mu m s^{-1}$ for long runs $(0.5-1~ \mu m)$ \cite{grosslipid}. From this, we estimate the switching rates to be in the range of $\sim (1-4) s^{-1}$. Assuming a step size for a single kinesin motor be $8 nm$ and using the measured values for average run velocity, we estimate the typical translocation rate $a$ to be $\sim (25-50) s^{-1}$, This would imply that the typical range of $Q$ would be between $3-25$.

Can we employ the same theoretical mean field strategy, as in the $Q=0$ limit ? Unfortunately we find that  the mean field predictions fail completely when $Q>0$, as is revealed by comparison with Monte Carlo simulations. Our `exact' analytic study in the $Q=0$ limit  is nevertheless useful in providing a conceptual basis to analyze the numerical simulations when $Q>0$.


In Fig.\,4 we show the density profile obtained for different values of $Q$, keeping the entry and exit rates of particles at the boundaries fixed. Fig.\,4(a) shows how for values of $Q$ close $0$, the MF prediction for densities matches well with the Monte-Carlo simulations. The parameter values shown in the figure correspond to a region in Low Density (LD) phase. On increasing $Q$, the bulk density increases continuously and starts deviating from the MF prediction. However the ratio of densities of the two species expressed in eq.(\ref{eq:ratio}) remains  unaltered with change in $Q$. The corresponding value of current in the system also deviates from the MF value and decreases continuously with the increase in Q. Fig.\,4(b) shows that at $Q \approx 0.84$, there is a first order transition, associated with a discontinuous change of density from a low density (LD) phase  to a high density (HD) phase. The current remains continuous across the phase transition. This non-equilibrium phase transition is unique, arising out of an interplay between translocation and switching dynamics in this driven system. However if the initial phase at $Q=0$ is either a HD phase or maximal current (MC) phase then no such discontinuous change of density is observed on increasing $Q$. Instead the bulk density increases continuously with increase in $Q$, saturating to a maximum value.

 Fig.\,4(a) shows how the value of the density in the HD phase is significantly higher than the MF prediction.    On further increasing $Q$, the density of $(+)$ species increases and asymptotically approaches a maximum value. Correspondingly as shown in Fig.\,5(a), the vacancy density reduces. 
With an increase of $Q$, the net current in the system approaches zero. Further, the steady state also becomes independent of the boundaries for higher values of $Q$. From the log-log plot it can inferred that the net current goes as, $J \sim 1/Q$, for large $Q$. 
This would imply that $J \sim k_{\pm}$; so that when the switching rates are much smaller than the translocation rates, then the net particle flux in the lattice is controlled by the switching process alone.

However, as far as we can tell from our numerical studies of this model, we do not observe any jamming phase transition at finite $Q$, upto the values of $Q = 500$ that we have investigated. Beyond that due to the smallness of values of the current, statistically reliable values of current is difficult to obtain numerically.
Figure 5(b) shows how $J_{+}$ reduces and system approaches jamming.  Although the average steady state profile is homogeneous, there is an increased tendency for the vacancies to cluster, as the system approaches jamming. 
 
It is interesting to compare the steady state properties of the current model with other closely related model. In particular, a two species model proposed in \cite{arndt}, with dynamic rules : $+ 0 \rightarrow 0 +$, $ 0-\rightarrow -0$ , $+ -\rightleftharpoons-+$, has some similarity with the dynamics of the model discussed here. It was found that for a closed ring, for a certain range of relative rates for the process  $+ -\rightleftharpoons-+$, there is a phase transition associated with the formation of three distinct domain which either move with constant drift velocity or remain in a pinned configuration, depending on the ratio of two species of particles. Apart from the nature of the boundary conditions, the crucial  difference between the two models is the aspect of Charge-Parity (CP) conservation -- while charge-parity is conserved in \cite{arndt}, it is explicitly broken in our case because of the switching dynamics. We would like to explore whether the  model that we discuss in this paper can also support such novel steady states, if we change the boundary conditions from open to closed.

Based only on our current numerical analysis it is difficult to assess whether this model has a finite $Q$ phase transition to a jammed state with zero particle current or simply a crossover  to jamming -- clearly this issue needs to addressed analytically.


\section{Conclusions}
\label{sec:conclusions} 

In this paper, we study a minimal lattice model which describes bidirectional transport of ``particles'' driven along a one dimensional track, as is observed in 
microtubule based, motor protein driven bidirectional transport of cargo vesicles, lipid bodies and organelles such as mitochondria.  This model can be described as
 a multi-species totally asymmetric exclusion process (TASEP)  with directional switching and which obeys overall particle number conservation in the bulk.  
It provides a simple framework for understanding  the interplay between the switching dynamics of individual particles and the collective movement of particles in 1-dimension.




 The model exhibits various interesting features such as current reversal of the net cargo, depending on the effective ratio of the switching rate $K_{+}$. This aspect of the model is consistent with flux reversals observed in the transport of pigment granules in melanophore cellular extracts, wherein changing the hormonal conditions leads to alterations in the (un)binding rates of the motors \cite{borisy}. The model predicts a
multicritical behaviour; the maximal current phase is absent for the entire physical range of $(\alpha_{+}-\beta_{+})$ phase plane beyond a critical value of $\alpha_{-}$. Switching rate induced phase transition is another interesting feature exhibited by this model : thus changes in the translocation rate or switching rates of the cargo using say biochemical methods, can lead to sharp changes in the average density profile of transported cargo. 

A possible extension of the present model is to go beyond local particle conservation by explicitly allowing  particles to exchange with the environment. Such a situation would typically arise when cargo vesicles moving along microtubules, detach from it and move along the surrounding actin meshwork for a short run time, before reattaching to the
same microtubule at a different location. Since the individual actin filaments in the actin meshwork are short and orientationally uncorrelated, the movement of the cargo vesicles in the actin meshwork is effectively diffusive. This suggests that the dynamics of cargo vesicles along a microtubule embedded in a confined isotropic actin meshwork
can be modeled by an appropriate non-conserved version of the current model.


\section{Acknowledgements}
We would like to thank A. Dhar for useful discussions and suggestions. MR thanks a grant from CEFIPRA 3504-2.


\begin{thebibliography}{}

\bibitem{borisy} V. I. Rodionov, A. G. Hope, T. M. Svitkina and G. G. Borisy, Curr.Biol. {\bf 8}, 165 (1998).

\bibitem{hollenbeck} R. L. Morris and P. J. Hollenbeck, J. Cell. Sc. {\bf 104}, 917 (1993).

\bibitem{welte} M. A. Welte, Curr.Biol. {\bf 14}, R525 (2004).

\bibitem{roop} V. Soppina, A.K. Rai, A. J. Ramaiya, P. Barak and R. Mallik, Proc. Nat. Acad. Sci. {\bf 106}, 19381 (2009).

\bibitem{lipo} M. J. I.  M\"uller, S. Klumpp and R. Lipowsky, Proc. Nat. Acad. Sci. {\bf 105}, 4609 (2008).

\bibitem{grosslipid} S.P. Gross, M. A. Welte and E. F. Wieschaus, J. Cell. Biol {\bf 156}, 715 (2002) 

\bibitem{cell} B. Alberts et al., {\it Molecular Biology of the Cell} ( Garland Science, New York, 2002).

\bibitem{howard} J. Howard, {\it Mechanics of Motor Proteins and the Cytoskeleton}, (Sinauer Associates, Massachusetts, 2001).

\bibitem{schutzrev} G M Schutz, J. Phys. A: Math. Gen. {\bf 36} R339 (2003)

\bibitem{nonequi2} Y. Kafri, E. Levine, D. Mukamel, G. M. Schutz and R. D. Wilmann, Phys. Rev. E. {\bf 68}, 035101(R) (2003).

\bibitem{privman}  V. Privman (ed.) in {\it Nonequilibrium Statistical Mechanics in One Dimension} (Cambridge University Press, Cambridge, 1997).

\bibitem{barma5} M. Barma and R. Ramaswamy in {\it Non-Linearity and Breakdown in Soft Condensed Matter} (Springer, Berlin, 1993).

\bibitem{kolo} A.B. Kolomeisky, G. M. Schutz, E. B. Kolomeisky  and J. P. Straley, J. Phys. A. {\bf 31}, 6911 (1998). 

\bibitem{freylet} A. Parmegianni, T. Franosch and E. Frey, Phys. Rev. Lett. {\bf 90} 086601 (2003). 

\bibitem{santen} R. Juhasz and L. Santen, J.Phys. A: Math. Gen. {\bf 37}, 3933 (2004).

\bibitem{thesis} S. Muhuri (Ph.D. Thesis, Jawaharlal Nehru University, New Delhi, 2006).

\bibitem{ignaepl} S. Muhuri and I. Pagonabarraga, EPL. {\bf 84}, 58009 (2008).

\bibitem{ignapre} S. Muhuri and I. Pagonabarraga, Phys. Rev. E. {\bf 82}, 021925 (2010).

\bibitem{gross} R. Mallik and S. P. Gross, Curr. Biol. {\bf 14}, R971 (2004). 

\bibitem{asepexact1} G. Schutz and E. Domany, J. Stat. Phys. {\bf 72}, 277 (1993).

\bibitem{asepexact2} B. Derrida, M. R. Hakim, V. Pasquier, J. Phys. A: Math. Gen. {\bf 26}, 1493 (1993).

\bibitem{arndt} P.F. Arndt and V. Rittenberg, J.Stat. Phys. {\bf 107}, 989 (2002)




















 


 









\end{thebibliography}
\end{document}